\documentclass[letterpaper,french,english,aps,showpacs,showkeys]{revtex4-1}
\usepackage[T1]{fontenc}
\usepackage[latin9]{inputenc}
\usepackage{color}
\usepackage{textcomp}
\usepackage{amsmath}
\usepackage{amssymb}
\usepackage{graphicx}

\makeatletter

\usepackage{babel}

\@ifundefined{showcaptionsetup}{}{%
 \PassOptionsToPackage{caption=false}{subfig}}
\usepackage{subfig}
\makeatother

\usepackage{babel}
\makeatletter
\addto\extrasfrench{%
   \providecommand{\fg}{\ifdim\lastskip>\z@\unskip\fi~\frqq}%
}

\makeatother
\begin{document}

\title{The statistical properties of the q-deformed Dirac oscillator in
one and two-dimensions}

\author{Abdelmalek Boumali}
\email{boumali.abdelmalek@gmail.com;boumali.abdelmalek@univ-tebessa.dz (Corresponding Author) }

\affiliation{Laboratoire de Physique Appliquée et Théorique, ~~\\
 \textcolor{black}{Université de Larbi-Tébessi}-Tébessa-, Algeria.}

\author{Hassan Hassanabadi}
\email{h.hasanabadi@shahroodu.ac.ir}

\affiliation{\textcolor{black}{Physics Department, Shahrood University of Technology,
}~~\\
 \textcolor{black}{Shahrood, Iran}}
\begin{abstract}
In this paper, we study the behavior of the eigenvalues of the one
and two dimensions of q-deformed Dirac oscillator. The eigensolutions
have been obtained by using a method based on the q-deformed creation
and annihilation operators in both dimensions. For a two-dimensional
case, we have used the complex formalism which reduced the problem
to the problem of one dimensional case. The influence of the q-numbers
on the eigenvalues has been well analyzed. Also, the connection between
the q-oscillator and a quantum optics is well established. Finally,
for very small deformation $\eta$, we have mentioned to existence
of well-known q-deformed version of Zitterbewegung in relativistic
quantum dynamics, and calculated the partition function and all thermal
quantities such as the free energy, total energy, entropy and specific
heat: here we consider only the case of a pure phase ($q=e^{i\eta}$).
The extension to the case of graphene has been discussed. 
\end{abstract}


\maketitle

\section{Introduction}

Quantum groups and quantum algebras have attracted much attention
of physicists and mathematicians during the last eight years . There
had been a great deal of interest in this field, especially after
the introduction of the q-deformed harmonic oscillator. Quantum groups
and quantum algebras have found unexpected applications in theoretical
physics \citep{1}. From the mathematical point of view they are q-deformations
of the universal enveloping algebras of the corresponding Lie algebras,
being also concrete examples of Hopf algebras. When the deformation
parameter q is set equal to 1, the usual Lie algebras are obtained.
The realization of the quantum algebra SU(2) in terms of the q-analogue
of the quantum harmonic oscillator \citep{2,3} has initiated much
work on this topic\citep{4,5,6}: Biedenharn and Macfarlane \citep{2,3}
have studied the q-deformed harmonic oscillator based on an algebra
of q-deformed creation and annihilation operators. They have found
the spectrum and eigenvalues of such a harmonic oscillator under the
assumption that there is a state with a lowest energy eigenvalue.
Recently, the theory of the q-deformed has become a topic of great
interest in the last few years, and it has been finding applications
in several branches of physics because of its possible applications
in a wide range of areas, such as a q-deformation of the harmonic
oscillator \citep{7}, a q-deformed Morse oscillator\citep{8}, a
classical and quantum q-deformed physical systems \citep{9}, Jaynes-Cummings
model and the deformed-oscillator algebra\citep{10}, q-deformed super-symmetric
quantum mechanics\textsc{ \citep{11}, }for some modified q-deformed
potentials \citep{12}, on the Thermo-statistic properties of a q-deformed
ideal Fermi gas \citep{13}, Q-Deformed Tamm-Dancoff oscillators \citep{14},
q-qeformed fermionic oscillator algebra and thermodynamics \citep{15},
and finally on the fermionic q deformation and its connection to thermal
effective mass of a quasi-particle \citep{16}.

The relativistic harmonic oscillator is one of the most important
quantum system, as it is one of the very few that can be solved exactly.
The Dirac relativistic oscillator (DO) interaction is an important
potential both for theory and application. It was for the first time
studied by Ito et al \citep{17}. They considered a Dirac equation
in which the momentum $\vec{p}$ is replaced by $\vec{p}-im\beta\omega\vec{r}$,
with $\vec{r}$ being the position vector, $m$ the mass of particle,
and $\omega$ the frequency of the oscillator. The interest in the
problem was revived by Moshinsky and Szczepaniak \citep{18}, who
gave it the name of Dirac oscillator (DO) because, in the non-relativistic
limit, it becomes a harmonic oscillator with a very strong spin-orbit
coupling term. Physically, it can be shown that the (DO) interaction
is a physical system, which can be interpreted as the interaction
of the anomalous magnetic moment with a linear electric field \citep{19,20}.
The electromagnetic potential associated with the DO has been found
by Benitez et al\citep{21}. The Dirac oscillator has attracted a
lot of interest both because it provides one of the examples of the
Dirac's equation exact solvability and because of its numerous physical
applications \citep{22,23,24,25,26,27}. We can note here that Franco-Villafane
et al \citep{28} have exposed the proposal of the first experimental
microwave realization of the one-dimensional (DO).

The q-deformed oscillator systems have attracted much attention and
have been considered in many papers (see Ref. \citep{29} and references
therein). The representation theory of the quantum algebras has led
to the development of q-deformed oscillator algebra. Since, there
have been an increasing interest in the study of physical systems
using q-oscillator algebra. It has found applications in several branches
of physics such as vibrational spectroscopy, nuclear physics, many
body theory and quantum optics. The q-analogue of the one-dimensional
non-relativistic harmonic oscillator has been studied by several authors
\citep{3,2,30,31}. Realizations of the quantum algebra $SU_{q}(1,1)$
via the one-dimensional q-harmonic oscillator were suggested by Chaichian
et al \citep{30} (see also Ref. \citep{32}). The representation
theory of quantum algebras with a single deformation parameter q,
has led to the development of tile now well-known q-deformed harmonic
oscillator algebra.

The extension of the non-relativistic q-harmonic oscillator to the
relativistic case, in the best of our knowledge, is not available
in the literature. In this context, and in order the overcome this
lack in the literature, the principal aim of this paper will be the
studied of q-deformed Dirac oscillator in one and two dimensions.
The concept of q-deformation is also applied to investigation of the
connection of q-deformed Dirac oscillator with quantum optics, and
the existence of the well-known Zitterbewegung in relativistic quantum
dynamics of the problem in question. In addition, we have evaluated
various thermodynamic quantities such as partition function, entropy
and internal energy. Such studies are expected to be relevant when
we want to extended them to the case of Graphene.

The structure of this paper is as follows: Sec. II is devoted to the
case of the standard q-harmonic oscillator. The extension to the q-deformed
Dirac oscillator will be treated in Sec. III. Different numerical
results about the thermal properties of q-deformed Dirac oscillator
are discussed in Sec. IV. Finally, Sec. V will be a conclusion.

\section{one-dimensional q-deformed standard harmonic oscillator: a review}

In the case of q-deformed harmonic oscillator, the creation and annihilation
operators $a+$ and $a$ satisfy the commutation relation \citep{34,35}
\begin{equation}
\left[a,a^{+}\right]_{q}=aa^{+}-q^{-1}a^{+}a=q^{N}\label{eq:1}
\end{equation}
where $N$ is the number operator, satisfying 
\begin{equation}
[N,a^{+}]=a^{+},[N,a]=-a.\label{2}
\end{equation}
The relevant Fock space is defined as 
\begin{equation}
a\left|0\right\rangle =0,\,\left|n\right\rangle =\frac{\left(a^{+}\right)^{n}}{\sqrt{\left[n\right]!}}\left|0\right\rangle ,\label{eq:3}
\end{equation}
where the q-factorial is defined as 
\begin{equation}
[n]!=[n][n-1]\cdot\cdot\cdot[1],\label{eq:4}
\end{equation}
and the q-numbers are defined by 
\begin{equation}
[n]=\frac{q^{n}-q^{-n}}{q-q^{-1}}.\label{eq:5}
\end{equation}
When q is real ($q=e^{\eta}$ ), the q-numbers take the form 
\begin{equation}
[k]=\frac{\sinh(\eta k)}{\sinh(\eta)}\label{eq:6}
\end{equation}
and when q is imaginary ($q=e^{i\eta}$ ), the q-numbers take the
form 
\begin{equation}
[k]=\frac{\sin(\eta k)}{\sin(\eta)}\label{eq:7}
\end{equation}
It is clear that in both cases $[k]\rightarrow k$ in the limit q
\textrightarrow{} 1.

The Hamiltonian of the q-deformed harmonic oscillator is 
\begin{equation}
H=\frac{P_{q}^{2}}{2m}+\frac{1}{2}m\omega^{2}Q_{q}^{2}\label{eq:8}
\end{equation}
where the q-momentum ($P_{q}$ ) and q-position ($Q_{q}$ ) operators
are directly written in terms of the q-boson operators $a$ and $a^{+}$
introduced above with 
\begin{equation}
P_{q}=i\sqrt{\frac{m\omega\hbar}{2}}\left(a-a^{+}\right),\label{eq:9}
\end{equation}
\begin{equation}
Q_{q}=\sqrt{\frac{\hbar\omega}{2m}}\left(a+a^{+}\right),\label{eq:10}
\end{equation}
Following this, we obtain 
\begin{equation}
H=\frac{\hbar\omega}{2}\left(aa^{+}+a^{+}a\right).\label{eq:11}
\end{equation}
The eigenvalues, in the Fock space defined above, are 
\begin{equation}
E_{n}=\frac{\hbar\omega}{2}\left(\left[n\right]+\left[n+1\right]\right).\label{eq:12}
\end{equation}
Hence the energy levels are no longer uniformly spaced as $q$ is
not equal to one. From the above Eq. (\ref{eq:12}), we find that
the q-deformed harmonic oscillator has a spectrum, given by 
\begin{equation}
E_{n}=\frac{\hbar\omega}{2}\frac{\sinh\left(\eta\left(n+\frac{1}{2}\right)\right)}{\sinh\left(\frac{\eta}{2}\right)},\label{eq:13}
\end{equation}
when $q$ is real, and by 
\begin{equation}
E_{n}=\frac{\hbar\omega}{2}\frac{\sin\left(\eta\left(n+\frac{1}{2}\right)\right)}{\sin\left(\frac{\eta}{2}\right)}\label{eq:14}
\end{equation}
when $q$ is complex. In both cases, when $q\rightarrow1\,\left(\eta\rightarrow0\right)$
the well-known relation 
\begin{equation}
E_{n}=\hbar\omega\left(n+\frac{1}{2}\right)\label{eq:15}
\end{equation}
is recover. According to the Eq. (\ref{eq:15}), one can see that
for $q$ real the energy eigenvalues increase more rapidly than the
ordinary case, in which the spectrum is equidistant, i.e. the spectrum
gets \textquotedbl{}expanded\textquotedbl{}. In contrast, when $q$
is a pure phase, the eigenvalues of the energy increase less rapidly
than the ordinary (equidistant) case, i.e. the spectrum is \textquotedbl{}compressed\textquotedbl{}
or squeezed.

In what follow, we treat the case of the one and two dimensional q-deformed
Dirac oscillator.

\section{Solutions of a Q-deformed Dirac oscillator}

\subsection{One-dimensional q-deformed Dirac oscillator}

The one-dimensional Dirac oscillator is 
\begin{equation}
\left\{ c\alpha_{x}\left(P_{q}-im\omega\beta Q_{q}\right)+\beta mc^{2}\right\} \psi_{D}=\epsilon\psi_{D},\label{eq:16}
\end{equation}
with $\psi_{D}=\left(\begin{array}{cc}
\psi_{1} & \psi_{2}\end{array}\right)^{T}$, $\alpha_{x}=\sigma_{x}$ and $\beta=\sigma_{z}$. In this case,
Eq. (\ref{eq:16}) becomes : 
\begin{equation}
H_{D}\psi_{D}=\epsilon\psi_{D},\label{eq:17}
\end{equation}
with 
\begin{equation}
H_{D}=\left(\begin{array}{cc}
mc^{2} & c\left(p_{x}+im\omega x\right)\\
c\left(p_{x}-im\omega x\right) & -mc^{2}
\end{array}\right)\label{eq:18}
\end{equation}
by introducing the usual annihilation and creation operators of the
q-deformed harmonic oscillator

\begin{equation}
P_{q}=i\sqrt{\frac{m\omega\hbar}{2}}\left(a-a^{+}\right),\label{eq:19}
\end{equation}
\begin{equation}
Q_{q}=\sqrt{\frac{\hbar\omega}{2m}}\left(a+a^{+}\right),\label{eq:20}
\end{equation}
this Hamiltonian transforms into 
\begin{equation}
H_{D}=\left(\begin{array}{cc}
mc^{2} & ga^{\dagger}\\
g^{*}a & -mc^{2}
\end{array}\right)\label{eq:21}
\end{equation}
with $g=imc^{2}\sqrt{2r}$, is the coupling strength between orbital
and spin degrees of freedom, and $r=\frac{\hbar\omega}{mc^{2}}=1$
is a parameter which controls the non-relativistic limit. It is an
important parameter that specifies the importance of relativistic
effects in the Dirac oscillator.

Writing that $\psi_{D}=\left(\left|n\right\rangle ,\left|n-1\right\rangle \right)^{T}$,
this equation can be solved algebraically. Following the above section,
when $q$ is real, the spectrum of energy is 
\begin{equation}
\epsilon_{n}=\pm mc^{2}\sqrt{1+2\frac{\sinh\left(\eta n\right)}{\sinh\left(\eta\right)}}.\label{eq:22}
\end{equation}
Now, if $q$ is complex, its becomes 
\begin{equation}
\bar{\epsilon}_{n}=\pm mc^{2}\sqrt{1+2\frac{\sin\left(\eta n\right)}{\sin\left(\eta\right)}},\label{eq:23}
\end{equation}
In both cases, when $q\rightarrow1\,\left(\eta\rightarrow0\right)$
the well-known relation 
\begin{equation}
\tilde{\epsilon}_{n}=\pm mc^{2}\sqrt{1+2n}\label{eq:24}
\end{equation}
is recover \citep{25}. The eigensolutions of a two-dimensional Dirac
oscillator, in both cases, can be written as 
\begin{equation}
\left|\psi\right\rangle =\left[\begin{array}{c}
\sqrt{\frac{E_{n}\pm mc^{2}}{2E_{nl}}}\left|n\right\rangle \\
\mp i\sqrt{\frac{E_{n}\mp mc^{2}}{2E_{nl}}}\left|n-1\right\rangle 
\end{array}\right].\label{eq:25}
\end{equation}
Using the creation and annihilation operators and the raising and
lowering operators $\sigma^{\pm}=\frac{1}{2}\left(\sigma^{x}\pm i\sigma^{y}\right)$
Dirac Spinor one can rewrite the previous equation as 
\begin{equation}
H_{1D}=g\left(\sigma^{\dagger}a+\sigma^{-}a^{\dagger}\right)+\Delta\sigma_{z}\label{eq:26}
\end{equation}
This Hamiltonian is exactly the JCM Hamiltonian in quantum optics
\citep{36}. Thus the one-dimensional Dirac oscillator maps exactly
onto the Jaynes-Cummings (JC), provided that one identifies the isospin
with the atomic system and the spatial degrees of freedom with the
cavity mode. Thus, as a result, we would like mentioned the relativistic
Hamiltonian of a q-deformed one-dimensional Dirac oscillator can be
mapped onto a q-deformed Jaynes-Cummings(JC).

\subsection{Two-dimensional q-deformed Dirac oscillator}

\subsubsection{Complex formalism}

In terms of complex coordinates and its complex conjugate \citep{37,38,39,40}
, we have 
\begin{equation}
z=x+iy,\,\bar{z}=x-iy,\label{eq:27}
\end{equation}
and 
\begin{equation}
\frac{\partial}{\partial z}=\frac{1}{2}\left(\frac{\partial}{\partial x}-i\frac{\partial}{\partial y}\right),\,\frac{\partial}{\partial\bar{z}}=\frac{1}{2}\left(\frac{\partial}{\partial x}+i\frac{\partial}{\partial y}\right).\label{eq:28}
\end{equation}
The operators momentum $p_{x}$ and $p_{y}$, in the Cartesian coordinates,
are defined by 
\begin{equation}
p_{x}=-i\hbar\frac{\partial}{\partial x},\,p_{y}=-i\hbar\frac{\partial}{\partial y}.\label{eq:29.}
\end{equation}
When we use $p_{z}=-i\hbar\frac{\partial}{\partial z}$, we get 
\begin{equation}
p_{z}=-i\hbar\frac{d}{dz}=\frac{1}{2}\left(p_{x}-ip_{y}\right),\label{eq:30}
\end{equation}
\begin{equation}
\bar{p}_{z}=-i\hbar\frac{d}{d\bar{z}}=\frac{1}{2}\left(p_{x}+ip_{y}\right),\label{eq:31}
\end{equation}
with $p_{z}=-\bar{p}_{z}$. These operators obey the basic commutation
relations 
\begin{equation}
\left[z,p_{z}\right]=\left[\bar{z},p_{\bar{z}}\right]=i\hbar,\,\left[z,p_{\bar{z}}\right]=\left[\bar{z},p_{z}\right]=0.\label{eq:32}
\end{equation}
The usual creation and annihilation operators, $a_{x}$ and $a_{y}$
with 
\begin{equation}
a_{x}=\sqrt{\frac{m\omega}{2\hbar}}x+i\frac{1}{\sqrt{2m\omega\hbar}}p_{x},\,a_{y}=\sqrt{\frac{m\omega}{2\hbar}}y+i\frac{1}{\sqrt{2m\omega\hbar}}p_{y},\label{eq:33}
\end{equation}
can be reformulated, in the formalism complex, as follows 
\begin{equation}
a_{z}=i\left(\frac{1}{\sqrt{m\omega\hbar}}\bar{p}_{z}-\frac{i}{2}\sqrt{\frac{m\omega}{\hbar}}z\right),\label{eq:34}
\end{equation}
\begin{equation}
\bar{a}_{z}=-i\left(\frac{1}{\sqrt{m\omega\hbar}}p_{z}+\frac{i}{2}\sqrt{\frac{m\omega}{\hbar}}\bar{z}\right).\label{eq:35}
\end{equation}
These operators, also, satisfy the habitual commutation relations
\begin{equation}
\left[a_{z},\bar{a}_{z}\right]=1,\,\left[a_{z},a_{z}\right]=0,\,\left[\bar{a}_{z},\bar{a}_{z}\right]=0.\label{eq:36}
\end{equation}
Now, In the case of q-deformed Dirac oscillator, the creation and
annihilation operators $\bar{a}_{z}$ and $a_{z}$ satisfy the commutation
relation 
\begin{equation}
\left[a_{z},\bar{a}_{z}\right]_{q}=a_{z}\bar{a}_{z}-q^{-1}\bar{a}_{z}a_{z}=q^{N}\label{eq:37}
\end{equation}
where $N$ is the number operator, satisfying 
\begin{equation}
[N,\bar{a}_{z}]=\bar{a}_{z},[N,a_{z}]=-a_{z}.\label{eq:38}
\end{equation}

\subsubsection{The solutions}

The two-dimensional Dirac oscillator is 
\begin{equation}
\left[c\sigma_{x}\left(p_{x}-im\omega x\right)+c\sigma_{y}\left(p_{y}-im\omega y\right)\right]\psi=\varepsilon\psi,\label{eq:39}
\end{equation}
with $\psi_{D}=\left(\begin{array}{cc}
\psi_{1} & \psi_{2}\end{array}\right)^{T}$, $\alpha_{x}=\sigma_{x}$ and $\beta=\sigma_{z}$. With the following
definitions of Dirac matrices, 
\begin{equation}
\alpha_{x}=\sigma_{x}=\left(\begin{array}{cc}
0 & 1\\
1 & 0
\end{array}\right),\,\alpha_{y}=\sigma_{y}=\left(\begin{array}{cc}
0 & -i\\
i & 0
\end{array}\right),\label{eq:40}
\end{equation}
Eq. (\ref{eq:14}) can be decoupled in a set of equations as follows
\begin{equation}
\varepsilon\left|\psi_{1}\right\rangle =c\left(p_{x}+im\omega x-ip_{y}+m\omega y\right)\left|\psi_{2}\right\rangle ,\label{eq:41}
\end{equation}
\begin{equation}
\varepsilon\left|\psi_{2}\right\rangle =c\left(p_{x}-im\omega x+ip_{y}+m\omega y\right)\left|\psi_{1}\right\rangle ,\label{eq:42}
\end{equation}
and so, Eq. (\ref{eq:13}) reads 
\begin{equation}
H_{D}=\left(\begin{array}{cc}
mc^{2} & c\left(p_{x}+im\omega x-ip_{y}+m\omega y\right)\\
c\left(p_{x}-im\omega x+ip_{y}+m\omega y\right) & -mc^{2}
\end{array}\right).\label{eq:43}
\end{equation}
This last form of Hamiltonian of Dirac can be written, in the complex
formalism, by 
\begin{equation}
H_{D}=\left(\begin{array}{cc}
mc^{2} & 2cp_{z}+im\omega c\bar{z}\\
2c\bar{p}_{z}-im\omega cz & -mc^{2}
\end{array}\right)=\left(\begin{array}{cc}
mc^{2} & 2g\bar{a}_{z}\\
2g^{*}a_{z} & -mc^{2}
\end{array}\right).\label{eq:44}
\end{equation}
Thus the problem is transformed to the one-dimensional case with a
complex variable $z$.

Now, Following Eqs. (\ref{eq:15}) and (\ref{eq:16}), the wave functions
$\psi_{1}$ and $\psi_{2}$ can be rewritten in the language of the
complex annihilation-creation operators as 
\begin{equation}
\left|\psi_{1}\right\rangle =\frac{g}{\varepsilon-mc^{2}}\bar{a}_{z}\left|\psi_{2}\right\rangle ,\label{eq:45}
\end{equation}
\begin{equation}
\left|\psi_{2}\right\rangle =\frac{g^{*}}{\varepsilon+mc^{2}}a_{z}\left|\psi_{1}\right\rangle .\label{eq:46}
\end{equation}
When we write the component $\left|\psi_{1}\right\rangle $ in terms
of the quanta bases, $\left|n\right\rangle =\frac{\left(a^{\dagger}\right)^{n}}{\sqrt{\left[n\right]!}}\left|0\right\rangle $,
these equations can be simultaneously diagonalized, and the energy
spectrum can be described by 
\begin{equation}
\varepsilon_{n}=\pm mc^{2}\sqrt{1+4\frac{\sinh\left[\eta n\right]}{\sinh\left(\eta\right)}}.\label{eq:47}
\end{equation}
when $q$ is real, and by 
\begin{equation}
\bar{\varepsilon}_{n}=\pm mc^{2}\sqrt{1+4\frac{\text{sin}\left[\eta n\right]}{\text{sin}\left(\eta\right)}},\label{eq:48}
\end{equation}
if $q$ is complex. In both cases, when $q\rightarrow1\,\left(\eta\rightarrow0\right)$
the well-known relation 
\begin{equation}
\tilde{\varepsilon}_{n}=\pm mc^{2}\sqrt{1+4n}\label{eq:49}
\end{equation}
is recovered. Our results are in a good agreement with those obtained
by Hatami et al \citep{41}.

According to last equation, the Dirac Hamiltonian can be written into
another form as

\begin{equation}
H_{2D}=g\left(\sigma^{\dagger}\bar{a}_{z}+\sigma^{-}a_{z}\right)+\Delta\sigma_{z},\label{eq:50}
\end{equation}
and it correspond to the q-deformed Anti-Jaynes-Cummings (AJC) model.
Here $\sigma^{\pm}=\frac{1}{2}\left(\sigma^{x}\pm i\sigma^{y}\right)$
are the spin arising and lowering operators, and $\Delta=mc^{2}$
is a detuning parameter. Before go further, we would like mentioned
the relativistic Hamiltonian of a q-deformed two-dimensional Dirac
oscillator can be mapped onto a couple of q-deformed Anti-Jayne-Cummings-
(AJC) which describe the interaction between the relativistic spin
and bosons.

We further observe that the Zitterbewegung frequency for the q-deformed
(2 + 1)-dimensional Dirac oscillator depends on the parameter of deformation
$\eta$. To show this we, first, start with the following eigensolutions
of a two-dimensional Dirac oscillator 
\begin{equation}
\left|\psi_{1,2}\right\rangle =\left[\begin{array}{c}
\sqrt{\frac{E_{n}\pm mc^{2}}{2E_{nl}}}\left|n\right\rangle \\
\mp i\sqrt{\frac{E_{n}\mp mc^{2}}{2E_{nl}}}\left|n-1\right\rangle 
\end{array}\right].\label{eq:51}
\end{equation}
Here, $E_{n}\equiv\epsilon_{n}$ $\left(\text{or}\,E_{n}\equiv\varepsilon_{n}\right)$
for q real (or q complex) respectively. The eigenstates can be expressed
transparently in terms of two-component Pauli spinors $\left|\chi_{\uparrow}\right\rangle $
and $\left|\chi_{\downarrow}\right\rangle $\citep{42,43,44,45} 
\begin{equation}
\left|\psi_{1}\right\rangle =\alpha_{n}\left|n\right\rangle \left|\chi_{\uparrow}\right\rangle -i\gamma_{n}\left|n-1\right\rangle \left|\chi_{\downarrow}\right\rangle ,\label{eq:52}
\end{equation}
\begin{equation}
\left|\psi_{2}\right\rangle =\gamma_{n}\left|n\right\rangle \left|\chi_{\uparrow}\right\rangle +i\alpha_{n}\left|n-1\right\rangle \left|\chi_{\downarrow}\right\rangle ,\label{eq:53}
\end{equation}
where $\alpha_{n}=\sqrt{\frac{\varepsilon_{n}+mc^{2}}{2\epsilon{}_{n}}}$
(or $\sqrt{\frac{\bar{\varepsilon}_{n}+mc^{2}}{2\varepsilon{}_{n}}}$)
and $\delta_{n}=\sqrt{\frac{\varepsilon_{n}-mc^{2}}{2\epsilon_{n}}}$
(or $\sqrt{\frac{\bar{\varepsilon}_{n}-mc^{2}}{2\varepsilon{}_{n}}}$)
are real. We can observe that the energy eigenstates present entanglement
between the orbital and spin degrees of freedom. The clarify this,
we start with some initial pure state at $t=0$, 
\begin{equation}
\left|\Psi\left(0\right)\right\rangle =\left|n-1\right\rangle \left|\chi_{\uparrow}\right\rangle =i\alpha_{n}\left|\psi_{1}\right\rangle -i\gamma_{n}\left|\psi_{2}\right\rangle ,\label{eq:54}
\end{equation}
This equation shows that the starting initial state is a superposition
of both the positive and negative energy solutions, which is the fundamental
ingredient that leads to Zitterbewegung in relativistic quantum dynamics.

The evolution of this initial state can be expressed as 
\begin{equation}
\left|\Psi\left(t\right)\right\rangle =i\gamma_{n}e^{-i\omega_{n}t}\left|\psi_{1}\right\rangle -i\alpha_{n}e^{i\omega_{n}t}\left|\psi_{2}\right\rangle \label{eq:55}
\end{equation}
where 
\begin{equation}
\omega_{n}=\frac{\epsilon_{n}}{\hbar}=\frac{mc^{2}}{\hbar}\sqrt{1+4\frac{\sinh\left(\eta n\right)}{\sinh\left(\eta\right)}}\label{eq:56}
\end{equation}
for $q$ real, and 
\begin{equation}
\omega_{n}=\frac{\bar{\epsilon}_{n}}{\hbar}=\frac{mc^{2}}{\hbar}\sqrt{1+4\frac{\text{sin}\left(\eta n\right)}{\text{sin}\left(\eta\right)}}\label{eq:57}
\end{equation}
for $q$ complex. In both cases, $\omega_{n}$ describes the frequency
of oscillations: the frequency oscillation between positive and negative
energy solutions.

If we consider very small deformation and neglect all terms proportional
to $\eta^{4}$, 
\begin{equation}
\omega_{n}=\frac{\epsilon_{n}}{\hbar}=\frac{mc^{2}}{\hbar}\sqrt{1+4\frac{\sinh\left[\eta n\right]}{\sinh\left(\eta\right)}}\approx\frac{mc^{2}}{\hbar}\sqrt{1+4n+\frac{2}{3}\eta^{2}n^{3}}\approx\omega_{DO}\left(1\pm\frac{1}{3}\frac{n^{3}}{\left(1+4n\right)}\eta^{2}\right),\label{eq:58}
\end{equation}
with $\omega_{DO}=\frac{mc^{2}}{\hbar}\sqrt{1+4n}$ is the frequency
of a two-dimensional Dirac oscillator without deformation, and the
sign $+$ denotes the case for q real and the sing $-$ for the case
of q complex.

Expression $\left|\Psi\left(t\right)\right\rangle $ in terms of two-component
$\left|\psi_{1}\right\rangle $ and $\left|\psi_{2}\right\rangle $,
and in the approximation of very small $\eta$, the final form will
be

\begin{align}
\left|\Psi\left(t\right)\right\rangle  & =\underbrace{\left|\Psi\left(t\right)\right\rangle _{\eta=0}}_{\text{without deformation}}+\eta^{2}\underbrace{\left|\Psi\left(t\right)\right\rangle _{\eta\neq0}}_{\text{en presence of deformation}},\label{eq:59}
\end{align}
where 
\begin{equation}
\left|\Psi\left(t\right)\right\rangle _{\eta=0}=\left(\cos\omega_{DO}t+\frac{i\sin\omega_{DO}t}{\sqrt{1+4n}}\right)\left|\psi_{1}\right\rangle +\sqrt{\frac{4n}{1+4n}}\sin\omega_{DO}t\left|\psi_{2}\right\rangle ,\label{eq:60}
\end{equation}
and 
\begin{align}
\left|\Psi\left(t\right)\right\rangle _{\eta\neq0} & =\left(\left[\frac{1}{3}\frac{n^{3}\omega_{DO}t}{\left(1+4n\right)}\sin\omega_{DO}t+\frac{i}{\sqrt{1+4n}}\left\{ \mp\frac{1}{3}\frac{n^{3}\omega_{DO}t}{\left(1+4n\right)}\left(\cos\omega_{DO}t+\sin\omega_{DO}t\right)\right\} \right]\right)\left|\psi_{1}\right\rangle +\nonumber \\
 & \left\{ \sqrt{\frac{4n}{1+4n}}\frac{1}{3}\frac{n^{3}\omega_{DO}t}{\left(1+4n\right)}\cos\omega_{DO}t-\sqrt{\frac{4n}{1+4n}}\left(\frac{n^{2}}{12}+\frac{1}{3}\frac{n^{3}}{\left(1+4n\right)}\right)\sin\omega_{DO}t\right\} \left|\psi_{2}\right\rangle .\label{eq:61}
\end{align}
The sign (-) for the case of q real, and (+) for the case of pure
phase. This equation shows the oscillatory behavior between the states
$\left|n\right\rangle \left|\chi_{\uparrow}\right\rangle $ and $\left|n-1\right\rangle \left|\chi_{\downarrow}\right\rangle $
which is exactly similar to atomic Rabi oscillations occurring in
the JC/AJC models. The q-deformed Rabi frequency is given by $\omega_{n}$
(Eq. (22)) for both cases.

\subsection{Discussions}

This section is devoted to study the influence of q-deformed algebra
on the eigenvalues of the Dirac oscillator in one and two dimensions.
This influence has been well established through the parameter $\eta$
with $q=e^{\eta}$.

\begin{figure}
\subfloat[q-deformed one-dimensional Dirac oscillator]{\includegraphics[scale=0.4]{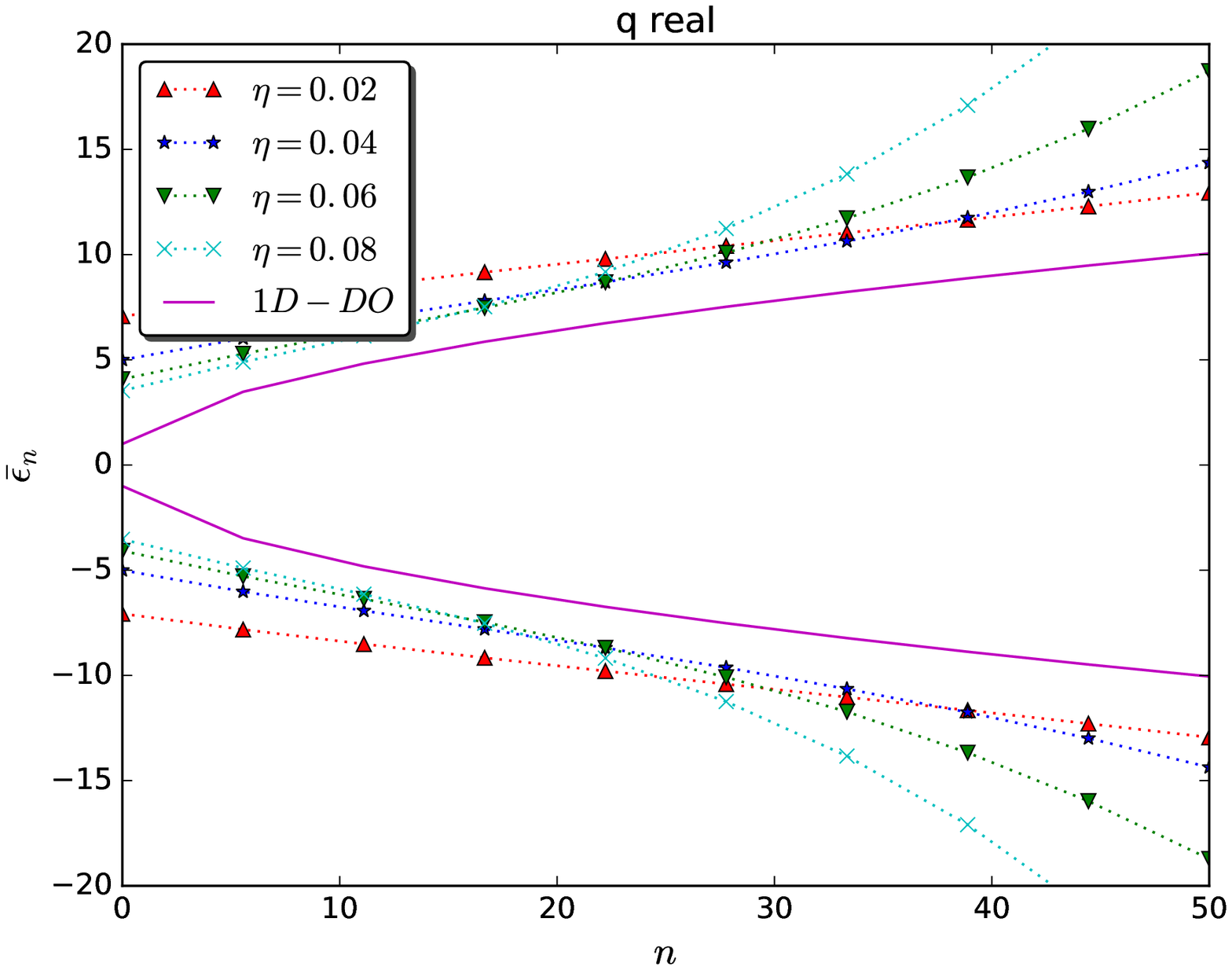} \includegraphics[scale=0.4]{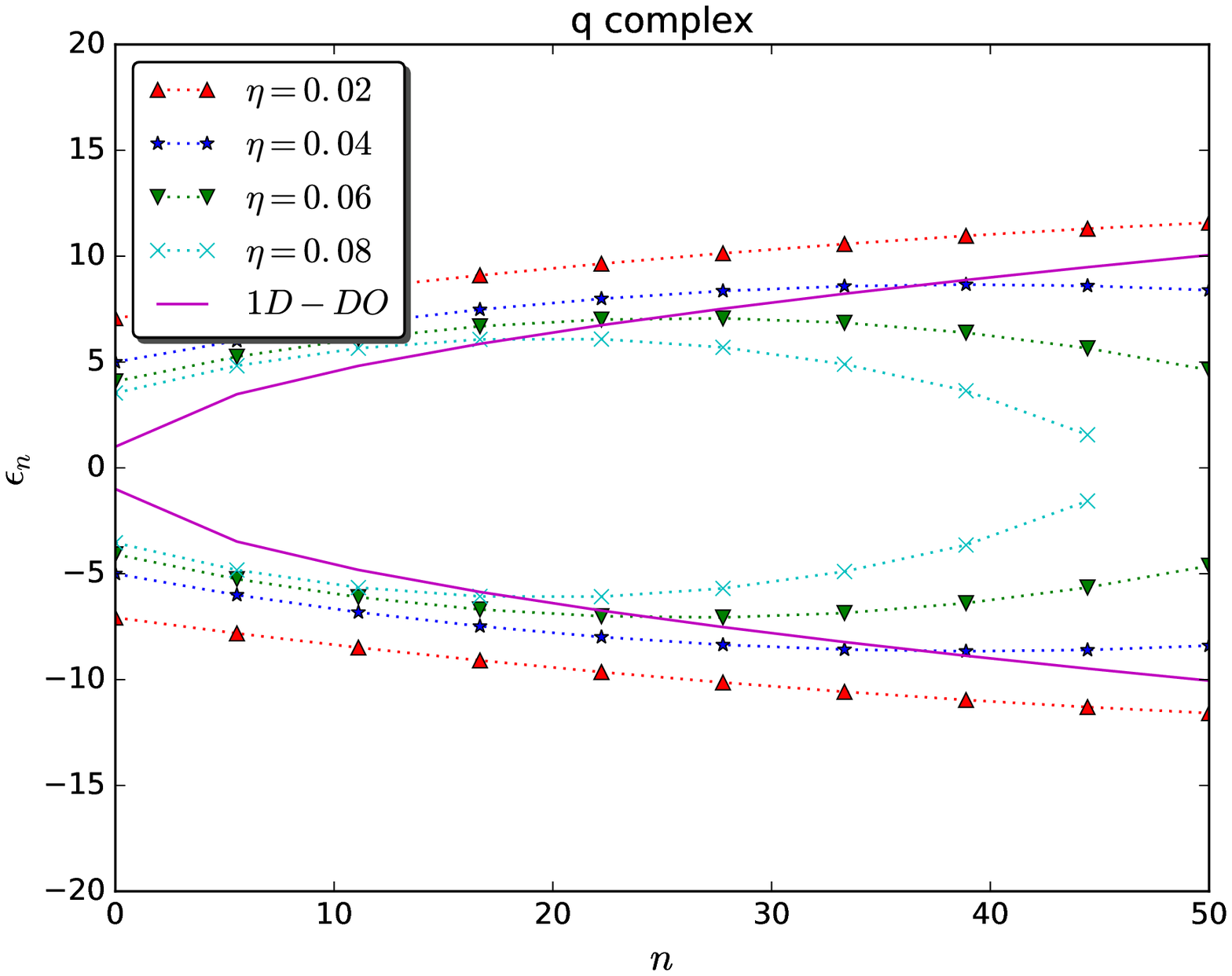}

}

\subfloat[q-deformed two-dimensional Dirac oscillator]{\includegraphics[scale=0.4]{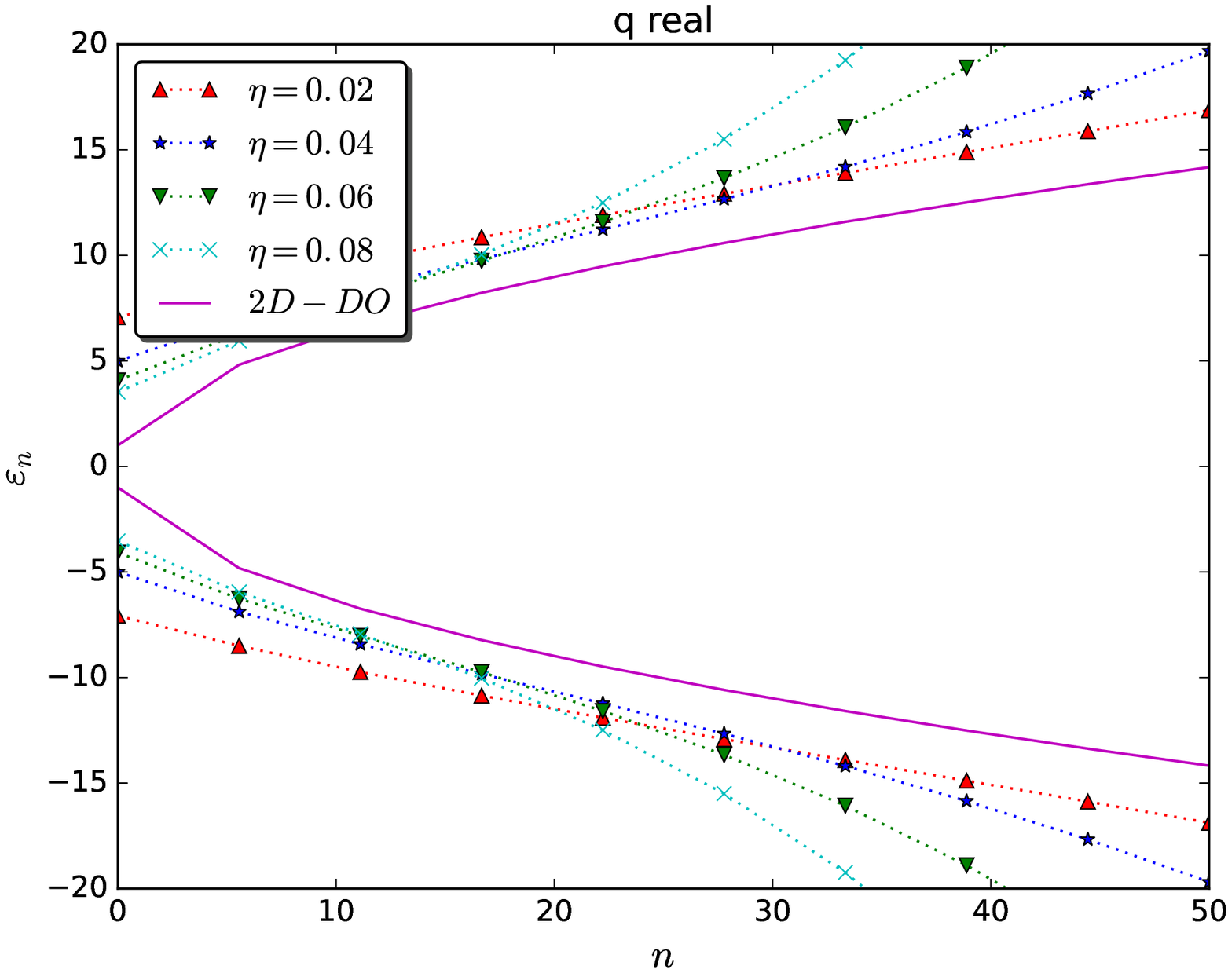} \includegraphics[scale=0.4]{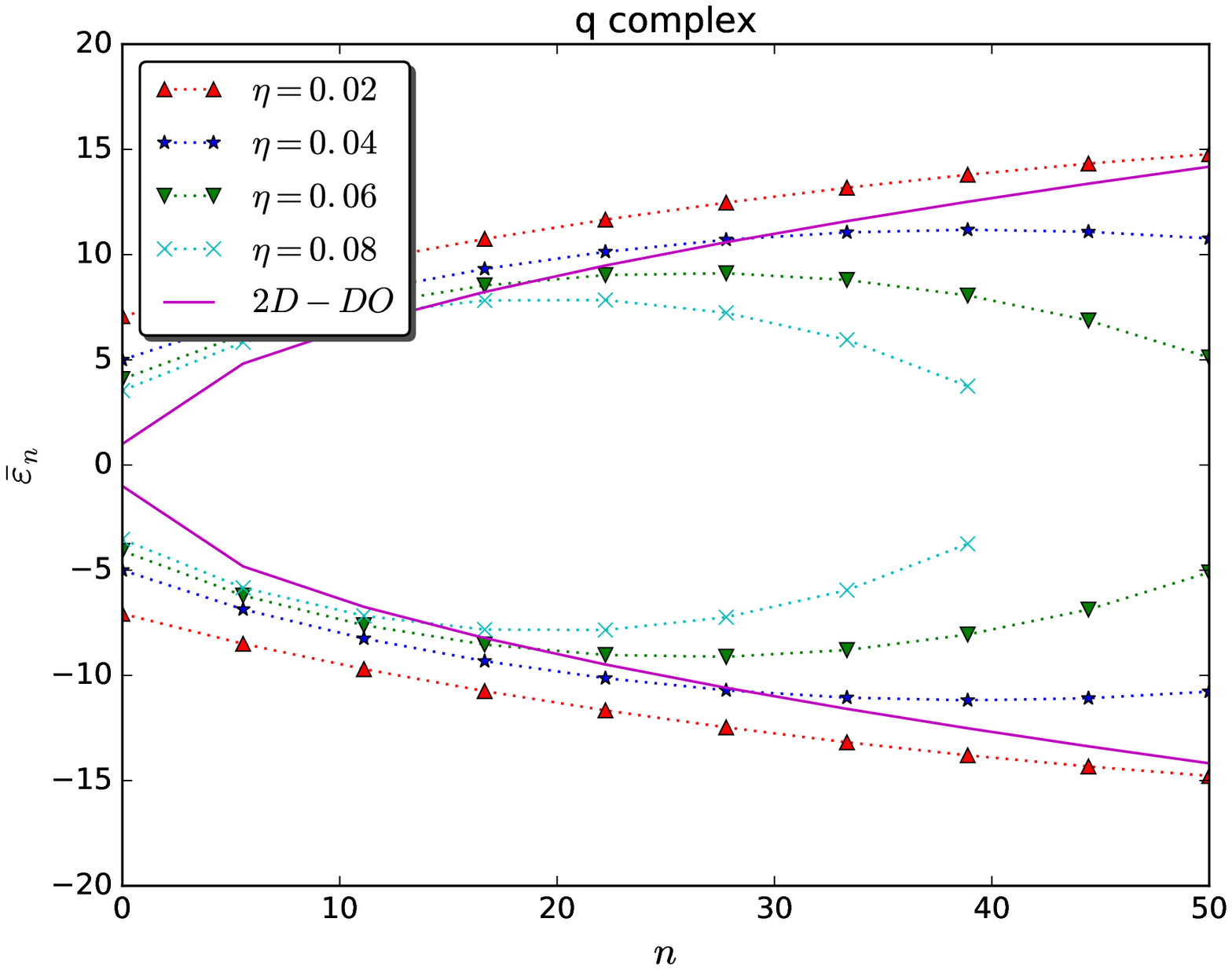}

}

\caption{\label{fig1}Spectrum of energy of one and two dimensions versus quantum
number for both q cases. }
\end{figure}

In Fig. \ref{fig1} 

, we present the eigenvalues of the q-deformed Dirac oscillator in
one and two dimensions versus the quantum number with different values
of parameter $\eta$ in both cases of q real and complex. In order
to argued this figure, we use the same explication used by Neskovic
and Urosevic \citep{47} in their study of the statistical properties
of quantum oscillator: thus, the energy levels of the q-oscillator
are not uniformly spaced for $q=1$. The behavior of the energy spectra
is completely different in the cases $q=e^{i\eta}$ and $q=e^{\eta}$.
When q is real$q=e^{i\eta}$, the separation between the levels increases
with the value of $n$ i.e, the spectrum is extended. On the other
hand, when q is a pure phase, the separation between all the levels
decreases with increasing $n$. i.e, the spectrum is squeeze.

\begin{figure}
\includegraphics[scale=0.4]{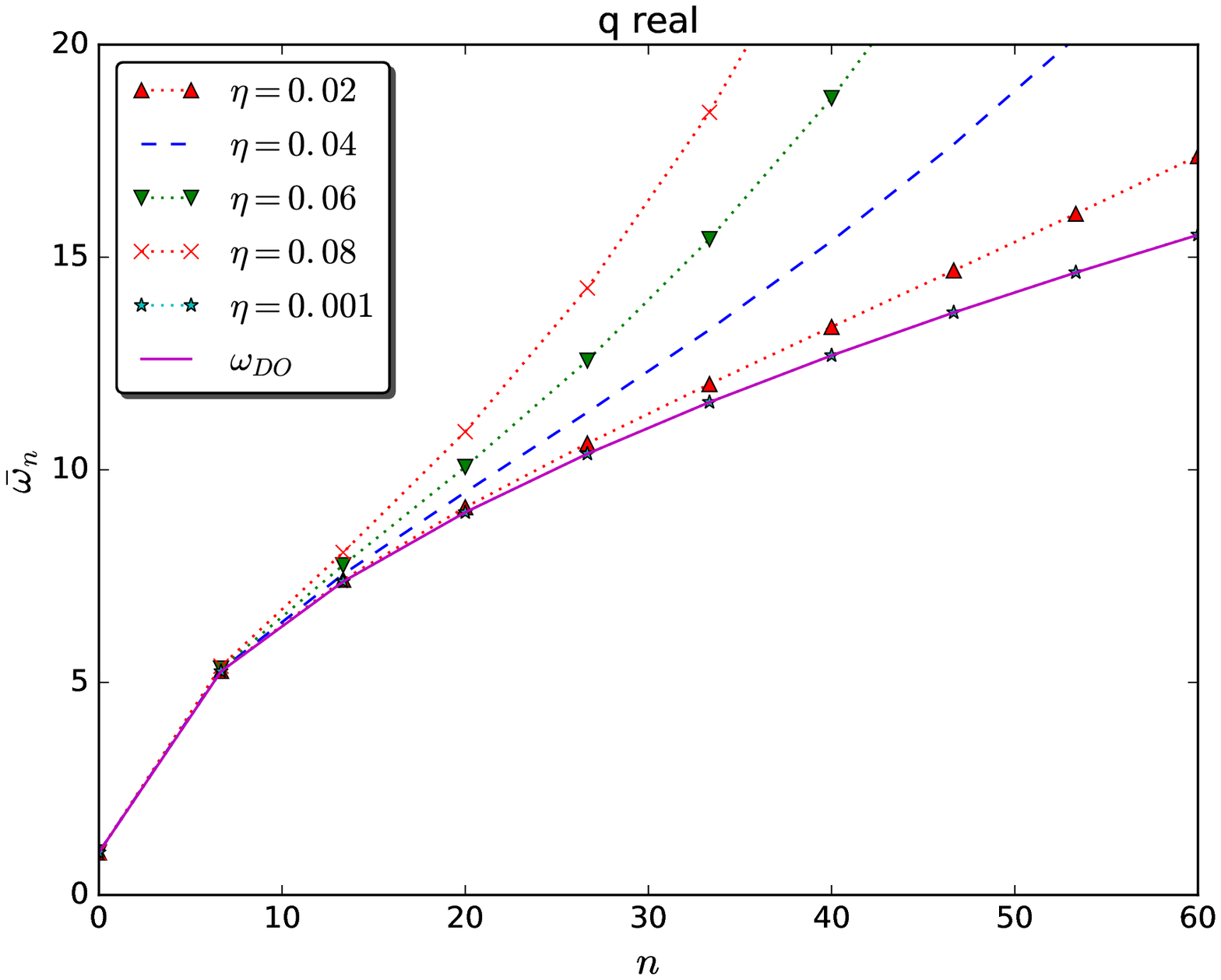} \includegraphics[scale=0.4]{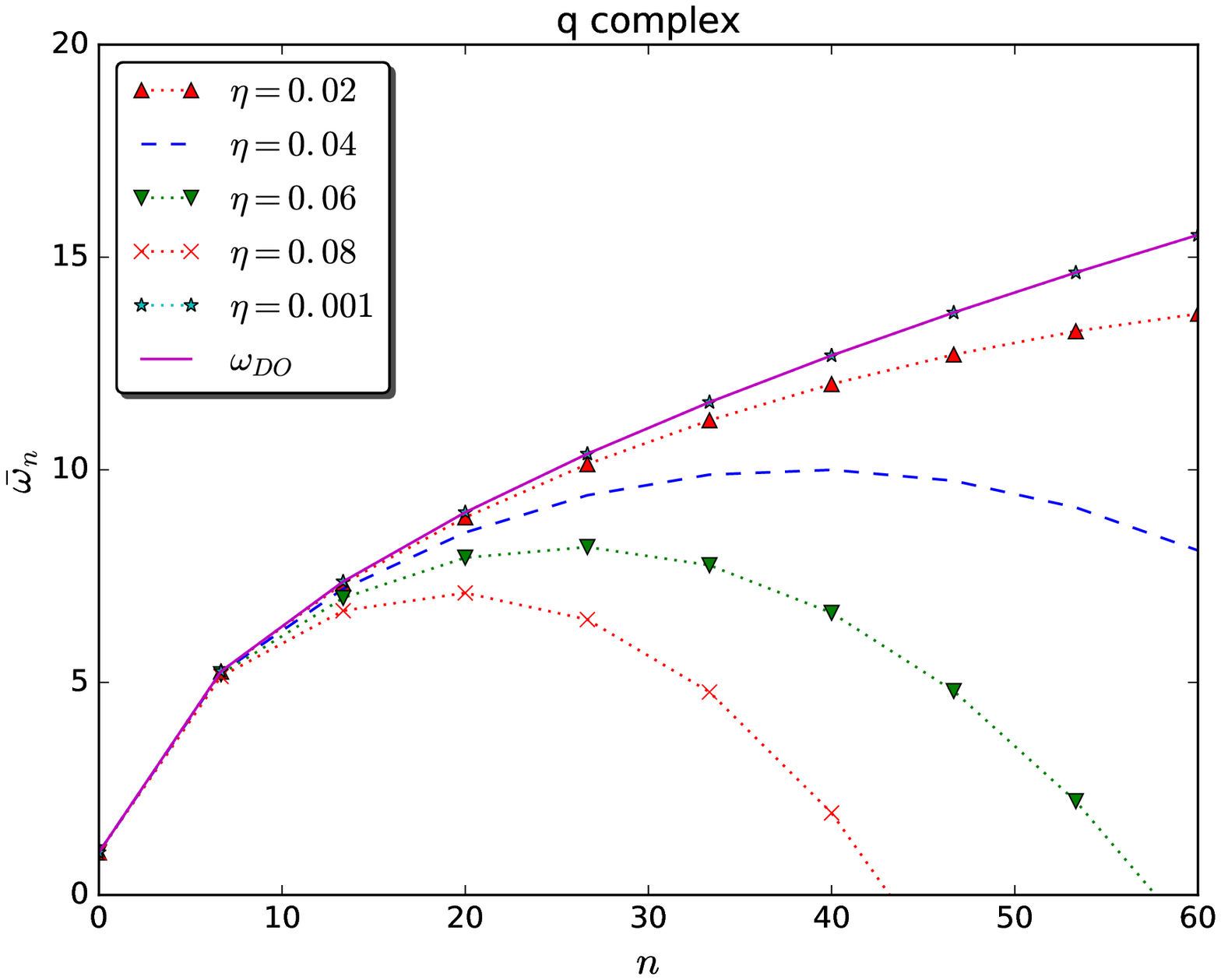}

\caption{\label{fig:2}Reduced frequency versus quantum number: here $\hbar=m=c=1$. }
\end{figure}

In Fig. \ref{fig:2}we present the frequency $\omega_{n}$ versus
a quantum number $n$ for both q real and complex in one and two-dimensions:
this frequency describes the oscillations between positive and negative
energy solutions. As consequence, we are in the case of well-known
Zitterbewegung in relativistic quantum dynamics. This phenomenon,
due to the interference of positive and negative energies, has never
been observed experimentally. The reason is that the amplitude of
these rapid oscillations lies below the Compton wavelength. The influence
of the deformation on this frequency is well established also. 

Finally, the exact connection of the q-deformed Dirac oscillator with
both Jaynes-Cummings (JC) and anti-Jaynes-Cummings (AJC) models has
been established.

\section{Thermal properties of q-deformed Dirac oscillator}

The theory of q-deformed statistics has become a topic of great interest
in the last few years because of its possible applications in a wide
range of areas, such as anyon physics, vertex models, quantum mechanics
in discontinuous space-time, quantum oscillator, and vibration of
polyatomic molecules, etc \citep{46}. In recent years, many researchers
have studied the q-deformed physical systems and have obtained a lot
of research progress. Among them, the statistical mechanics of q-deformed
quantum oscillator studied by Neskovic and Urosevic \citep{47}. Using
the boson realization of q-oscillator algebra and taking q to be real,
they have calculated the partition function Z and thermodynamic potentials
such as free energy F, entropy S and internal energy U for a Slightly
Deformed Oscillator.

The probability of finding a system in a state with energy $E_{n}$
is given by 
\begin{equation}
P_{n}=\frac{e^{-\frac{E_{n}}{k_{B}T}}}{Z},\label{eq:62}
\end{equation}
here $Z=\text{tr}e^{-\frac{H}{k_{B}T}}=\sum e^{-\beta E_{n}}$ is
the partition function. Our main object is to obtain this partition
function for small deformation $\beta$: in this case the summation
appears in $Z$ can be easily performed. Before doing so, lets rewrite
the form of energy in a more convenient form. Starting with the following
equation 
\begin{equation}
\epsilon_{n}=\pm mc^{2}\sqrt{1+a\frac{\sinh\left(\eta n\right)}{\sinh\left(\eta\right)}},\label{eq:63}
\end{equation}
for q real and 
\begin{equation}
\bar{\epsilon}_{n}=\pm mc^{2}\sqrt{1+a\frac{\text{sin}\left(\eta n\right)}{\text{sin}\left(\eta\right)}}\label{eq:64}
\end{equation}
for q complex. Here $a=2$ for one-dimensional ($4$ for a two-dimensional
case). Now, in order to extract these properties of our q-oscillator,
we will only restrict ourselves to stationary states of positive energy.
The Dirac oscillator possesses an exact Foldy\textendash Wouthuysen
transformation (FWT): so, the positive- and negative-energy solutions
never mix. Following this, we only consider the positive part of energy.

As $\frac{\sinh\left(\eta n\right)}{\sinh\left(\eta\right)}$ is even
as a function of $\eta$, so that the same property has the energy
and all quantities derived from it. Now, we will consider very small
deformation and neglect all terms proportional to $\eta^{4}$. In
this case, we have 
\begin{align}
\epsilon_{n} & \simeq\sqrt{1+an}\left(1+\frac{an^{3}}{12\left(1+an\right)}\eta^{2}\right)\label{eq:65}
\end{align}
With the same argument, the energy spectrum of the q complex case
can be written as 
\begin{equation}
\bar{\epsilon}=\backsimeq\sqrt{1+an}\left(1-\frac{an^{3}}{12\left(1+an\right)}\eta^{2}\right).\label{eq:66}
\end{equation}
A both equations can be written in a compact form as 
\begin{equation}
\xi_{n}=\sqrt{1+an}\left(1\pm\frac{an^{3}}{12\left(1+an\right)}\eta^{2}\right).\label{eq:67}
\end{equation}
or 
\begin{equation}
\xi_{n}=\xi_{n0}+\xi_{n\pm},\label{eq:68}
\end{equation}
with 
\begin{equation}
\xi_{n0}=\sqrt{1+an},\label{eq:69}
\end{equation}
is the reduced spectrum of energy of the ordinary Dirac oscillator,
and 
\begin{equation}
\xi_{n\pm}=\pm\frac{an^{3}}{12\sqrt{1+an}}\eta^{2}\label{eq:70}
\end{equation}
is the correction on the energy when the deformation exists.

In this case, the partition function is 
\begin{equation}
Z=\sum_{n=0}^{\infty}e^{-\frac{\xi_{n\pm}}{\tau}}\backsimeq\sum_{n=0}^{\infty}e^{-\frac{\sqrt{1+an}}{\tau}}\left(1\pm\frac{an^{3}}{12\tau\sqrt{1+an}}\eta^{2}\right)=Z_{0}+Z_{1}\label{eq:71}
\end{equation}
with $\tau=\frac{k_{B}T}{mc^{2}}$ and 
\begin{equation}
Z_{0}=\sum_{n=0}^{\infty}e^{-\frac{\sqrt{1+an}}{\tau}},\label{eq:72}
\end{equation}
and 
\begin{equation}
Z_{1}=\pm\frac{a\eta^{2}}{12\tau}\sum_{n=0}^{\infty}\frac{n^{3}}{\sqrt{1+an}}e^{-\frac{\sqrt{1+an}}{\tau}}.\label{eq:73}
\end{equation}
In order the evaluate the first term, $Z_{0}$, we use a method based
on Zeta function: so, by using the formula\citep{48}\citep{48} 
\begin{equation}
e^{-x}=\frac{1}{2\pi i}\int_{C}dsx^{-s}\Gamma\left(s\right),\label{eq:74}
\end{equation}
the sum in Eq. (\ref{eq:18}) is transformed into 
\begin{equation}
\sum_{n}e^{-\frac{\gamma}{\tau}\sqrt{\alpha+n}}=\frac{1}{2\pi i}\int_{C}ds\left(\frac{\gamma}{\tau}\right)^{-s}\sum_{n}\left(n+\alpha\right)^{-\frac{s}{2}}\Gamma\left(s\right)=\frac{1}{2\pi i}\int_{C}ds\left(\frac{\gamma}{\tau}\right)^{-s}\zeta_{H}\left(\frac{s}{2},\alpha\right)\Gamma\left(s\right),\label{eq:75}
\end{equation}
with $x=\frac{\gamma}{\tau}\sqrt{\alpha+n},\,\gamma=\sqrt{a},\,\alpha=\frac{1}{a}$.
Here, $\Gamma\left(s\right)$ and $\zeta_{H}\left(\frac{s}{2},\alpha\right)$
are respectively the Euler and Hurwitz zeta function. Applying the
residues theorem, for the two poles $s=0$ and $s=2$, the desired
partition function is written down in terms of the Hurwitz zeta function
as follows: 
\begin{equation}
Z_{0}\left(\tau\right)=\frac{\tau^{2}}{a}+\zeta_{H}\left(0,\frac{1}{a}\right).\label{eq:76}
\end{equation}
Now, the second term can be evaluated by using the Euler\textendash MacLaurin
formula; starting by the following equation 
\begin{equation}
Z_{1}=\pm\frac{a\eta^{2}}{12\tau}\sum_{n=0}^{\infty}\frac{n^{3}}{\sqrt{1+an}}e^{-\frac{\sqrt{1+an}}{\tau}}\label{eq:77}
\end{equation}
and according this approach, the sum transforms to the integral as
follows 
\begin{equation}
\sum_{x=0}^{\infty}f\left(x\right)=\frac{1}{2}f\left(0\right)+\int_{0}^{\infty}f\left(x\right)dx-\sum_{p=1}^{\infty}\frac{B_{2p}}{\left(2p\right)!}f^{\left(2p-1\right)}\left(0\right),\label{eq:78}
\end{equation}
Here $f\left(x\right)=\frac{x^{3}}{\sqrt{1+ax}}e^{-\frac{\sqrt{1+ax}}{\tau}}$,
$B_{2p}$ are the Bernoulli numbers, $f^{\left(2p-1\right)}$ is the
derivative of order (2 p \textminus{} 1). Up to $p=1$, the final
form of $Z_{1}$ term is: 
\begin{equation}
Z_{I}\left(\tau,\eta\right)=\int_{0}^{\infty}f\left(x\right)dx=b\eta^{2}e^{-\frac{1}{\tau}}\left(15\tau^{6}+15\tau^{5}+6\tau^{4}+\tau^{3}\right),\label{eq:79}
\end{equation}
with $b=6e^{-\frac{1}{\tau}}$ when $a=2$, and $b=\frac{e^{-\frac{1}{\tau}}}{8}$
when $a=4$. Finally, the compact final form of the q-deformed partition
function of the Dirac oscillator in one and two-dimensions for both
cases of q is 
\begin{equation}
Z_{q}\left(\tau,\eta\right)=\frac{\tau^{2}}{2}+\zeta_{H}\left(0,\frac{1}{a}\right)\pm\eta^{2}b\left(15\tau^{6}+15\tau^{5}+6\tau^{4}+\tau^{3}\right).\label{eq:82}
\end{equation}
Here, the sign (-) describes the partition function in the case where
q is real, and (+) the case of q complex

Via Eq. (\ref{eq:40}), the determination of all thermal properties,
such as the free energy, the entropy, total energy and the specific
heat, can be obtained through the numerical partition function $Z\left(\tau\right)$
via the following relations \citep{19}\textcolor{black}{{} 
\begin{equation}
F=-\tau\ln\left(Z\right),\,U=\tau^{2}\frac{\partial\ln\left(Z\right)}{\partial\tau},\label{eq:83}
\end{equation}
\begin{equation}
\frac{S}{k_{B}}=\ln\left(Z\right)+\tau\frac{\partial\ln\left(Z\right)}{\partial\tau},\,\frac{C}{k_{B}}=2\tau\frac{\partial\ln\left(Z\right)}{\partial\tau}+\tau^{2}\frac{\partial^{2}\ln\left(Z\right)}{\partial\tau^{2}}.\label{eq:84}
\end{equation}
}

\subsection{Numerical results}

\begin{figure}
\includegraphics[scale=0.4]{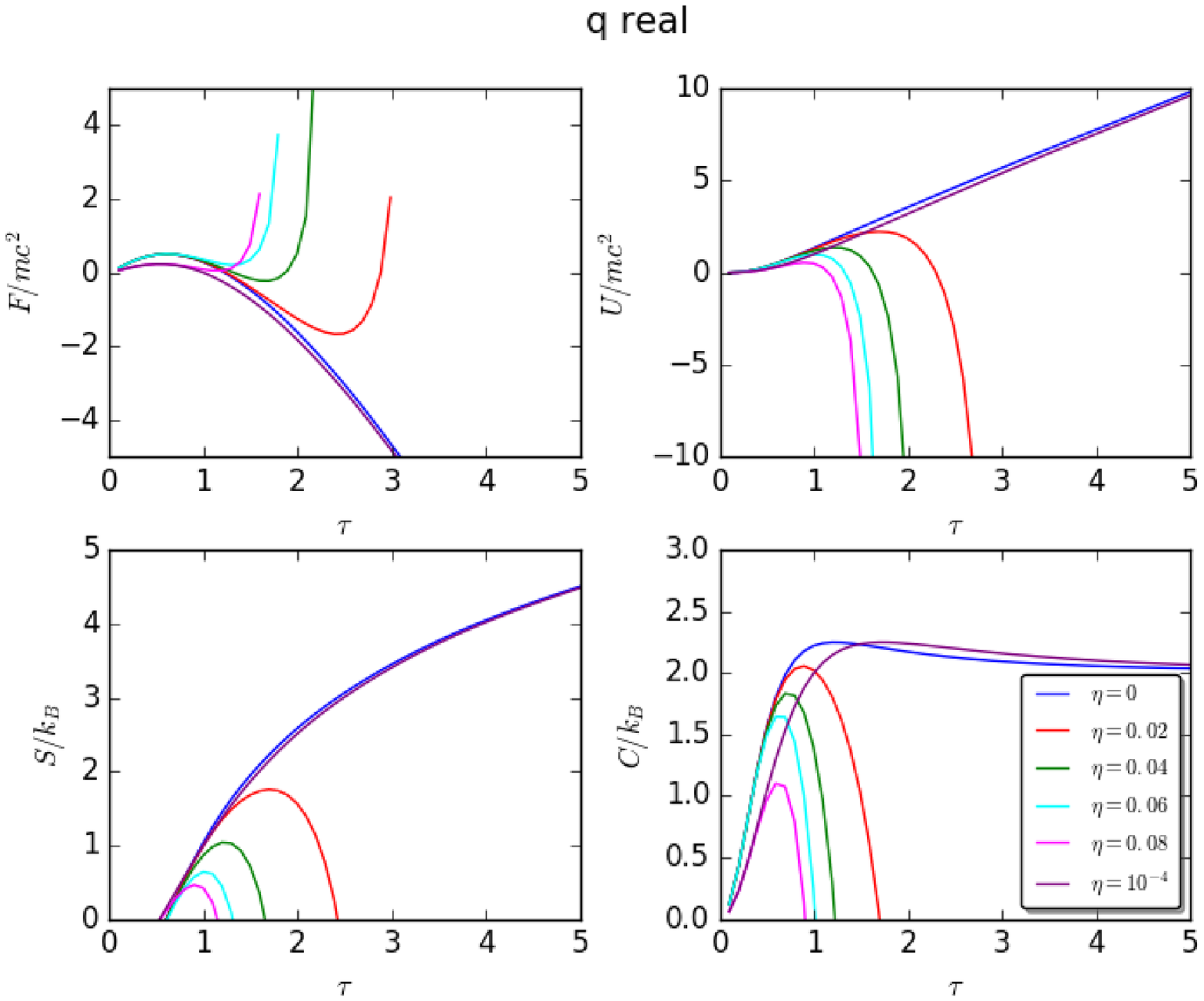} \includegraphics[scale=0.4]{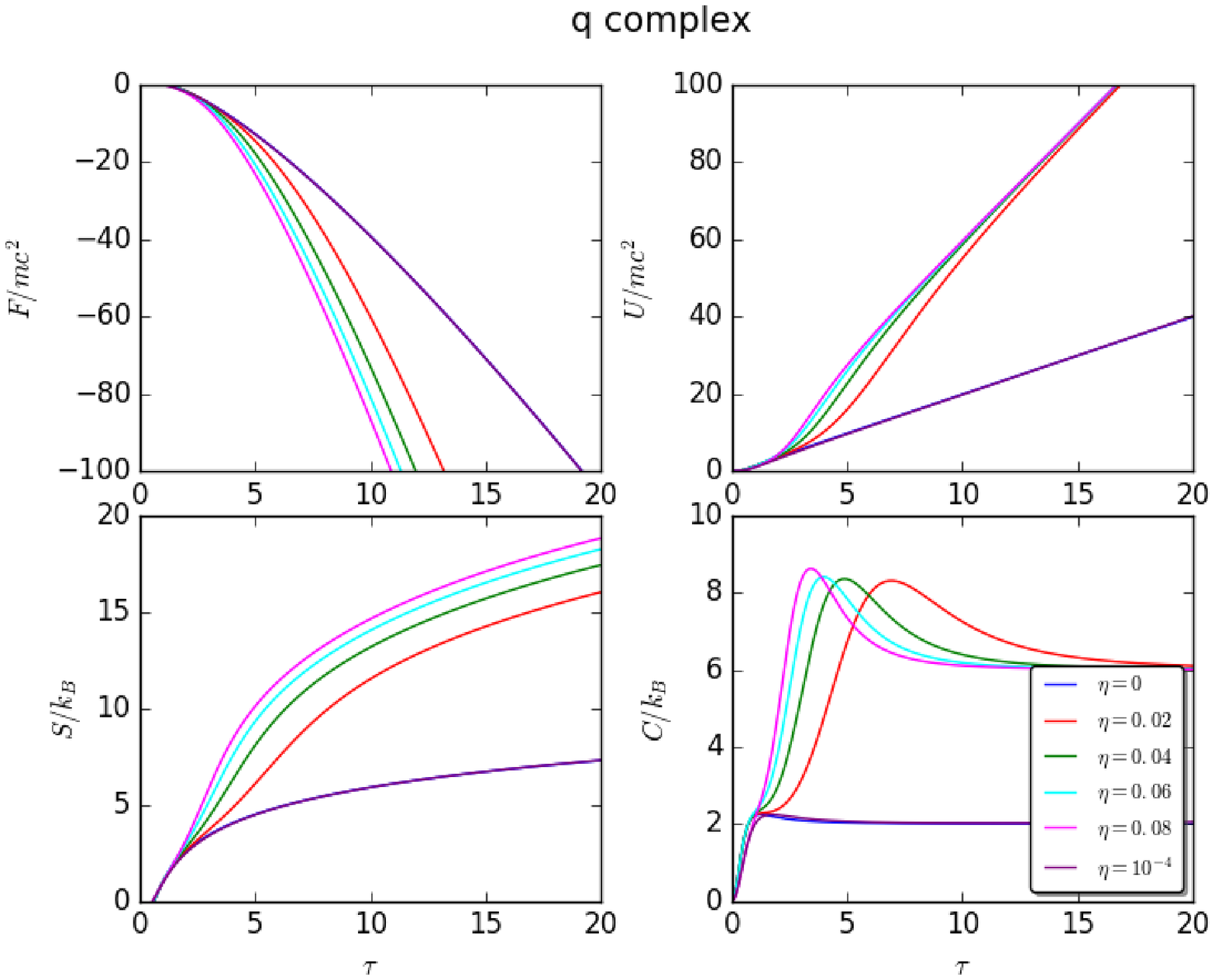}

\caption{\label{fig:3}Thermal properties of a q-deformed Dirac oscillator
in one and two dimensions}
\end{figure}

The Figure. \ref{fig:3} 

shows the thermal properties of a one and two-dimensional q-deformed
Dirac oscillator in both cases of q. from this figure, we can confirm
that the deformation plays a significant role on these properties,
and the effect of the parameters is very important on the thermodynamic
properties. Also, the behavior of these quantities are completely
different in the cases $q=e^{i\eta}$ and $q=e^{\eta}$: When q is
a pure phase, the behavior of different thermal quantities have a
similar comportment as in the case of non-deformed Dirac oscillator
in both one -and two dimensions \citep{48,49}. On the other hand
if q is real, these quantities show a strange behavior. This situation
is closely related to the nature of spectrum: if q is real, the spectrum
is extended which is the cause of the strange behavior of the thermodynamics
quantities, contrary to the case where q is complex where the spectrum
is squeeze: in this case we obtain the same form of all curves in
these quantities. In what follow, in order to compare our results
with those obtained in literature, we focus on the case of pure phase.

We should mention that, in all figures, we have used dimensionless
quantities, and the temperature range is taken from $10^{8}\text{K}\,\text{to}\,10^{14}\text{K}$.
These values give an order of the oscillator frequency about $10^{20}\text{Hz}$
similar to that of Zitterbewegung frequency in the DO, which has so
far been experimentally inaccessible. For the asymptotic limits, as
are shown by the figures of the specific heat in the presence of deformation,
all curves coincide, and reach the fixed value $C=6\text{\ensuremath{k_{B}}}$
three times greater compared to the case of non-deformed Dirac oscillator
in one and two dimensions.

As an application, we can extend our calculations to the case of graphene:
Graphene is a two-dimensional configuration of carbon atoms organized
in a hexagonal honeycomb structure. The electronic properties of graphene
are exceptionally novel. For instance, the low-energy quasi-particles
in Graphene behave as massless chiral Dirac fermions, which has led
to the experimental observation of many interesting effects similar
to those predicted in the relativistic regime. In the recent study
\citep{49}, the author has shown, by using an approach based on the
effective mass, that the model of a two-dimensional Dirac oscillator
can be used to describe the thermodynamic properties of Graphene under
an uniform magnetic field. By using the formalism of the creation
and annihilation operators in complex formalism, he arrives at the
following spectrum of energy 
\begin{equation}
\varsigma_{n}^{\pm}=\pm\sqrt{2}\frac{\hbar\tilde{c}}{l_{B}}\sqrt{n},\label{eq:85}
\end{equation}
with $l_{B}=\sqrt{\frac{\hbar}{eB}}$ is the so-called magnetic length
and $\tilde{c}$ is the Fermi velocity of electrons in the graphene.
The form of this spectrum of energy is in good agreement with the
form of the case of graphene( See Ref \citep{50}). This form of spectrum
of energy in the presence of a deformation $q$ can be written in
a pure phase by 
\begin{equation}
\bar{\zeta}_{n}^{\pm}=\pm\sqrt{2}\frac{\hbar\tilde{c}}{l_{B}}\sqrt{\frac{\sin\left(\eta n\right)}{\sin\left(\eta\right)}}.\label{eq:86}
\end{equation}
In the approximation of very small deformation, we have 
\begin{equation}
\frac{\zeta_{qn}}{a1}=\varsigma_{n}^{+}\left(1+\frac{n^{2}}{12}\eta^{2}\right),\label{eq:87}
\end{equation}
with $a1=\frac{\sqrt{2}\hbar\tilde{c}}{l_{B}},$ and consequently,
the final partition function of the q-deformed version of graphene
is 
\begin{equation}
Z_{q}^{'}\left(\tau,\eta\right)=Z_{0}^{'}+Z_{1}^{'},\label{eq:88}
\end{equation}
with 
\[
Z_{0}^{'}=\tau^{2}+\frac{1}{2},
\]
according to the Ref.\citep{49}, and 
\[
Z_{1}^{'}=\frac{\eta^{2}}{12\tau}\sum_{n=0}^{\infty}n^{2}e^{-\frac{\sqrt{n}}{\tau}}.
\]
Here $\tau=\frac{k_{B}l_{B}T}{\sqrt{2}\hbar\tilde{c}}.$ In order
to evaluate $Z_{1}$, we use the Euler\textendash MacLaurin formula:
so we obtain 
\[
Z_{1}^{'}=20\eta^{2}\tau^{5}.
\]
Thus, the final form of partition function is 
\[
Z_{q}^{'}\left(\tau,\eta\right)=\tau^{2}+\frac{1}{2}+20\tau^{5}\eta^{2}.
\]
Via equation (41), the determination of all thermal properties, such
as the free energy, the entropy, total energy and the specific heat,
can be obtained through the numerical partition function $Z_{q}^{'}\left(\tau,\eta\right)$
via the following relations\textcolor{black}{{} 
\begin{equation}
\frac{F}{a1}=-\frac{1}{\bar{\beta}}\ln\left(Z\right)=-\tau\ln\left(Z\right),\frac{U}{a1}=-\frac{\partial\ln\left(Z\right)}{\partial\bar{\beta}}=\tau^{2}\frac{\partial\ln\left(Z\right)}{\partial\tau},\label{eq:41-1}
\end{equation}
\begin{equation}
\frac{S}{k_{B}}=\bar{\beta}^{2}\frac{\partial\left(\frac{F}{a}\right)}{\partial\bar{\beta}}=\ln\left(Z\right)+\tau\frac{\partial\ln\left(Z\right)}{\partial\tau},\,\frac{C}{k_{B}}=-\bar{\beta}^{2}\frac{\partial\left(\frac{U}{a}\right)}{\partial\bar{\beta}}=2\tau\frac{\partial\ln\left(Z\right)}{\partial\tau}+\tau^{2}\frac{\partial^{2}\ln\left(Z\right)}{\partial\tau^{2}}.\label{eq:42-1}
\end{equation}
}

\begin{figure}
\includegraphics[scale=0.6]{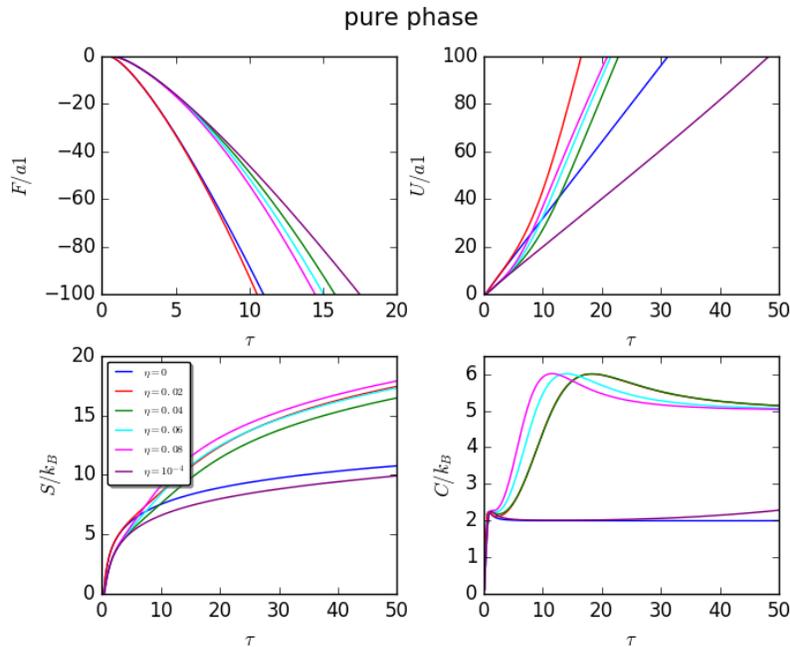}

\caption{\label{fig:4}Thermal properties of graphene in its q-deformed version}
\end{figure}

The thermodynamic quantities are, respectively, plotted in Figure.
\ref{fig:4}. From this figure, we observe that the behavior of the
specific heat in the asymptotic regions is greater that in the case
of graphene: here the limit is $5k_{B}$ . In addition, and as in
the non-deformed case of graphene, we can argued this situation by
saying that these limits follow the Dulong\textendash Petit law for
an ultra-relativistic ideal gas 

\section{Conclusion}

In this paper, after a brief preliminary around q-deformed oscillator,
we studied the Dirac oscillator in this deformation formalism. We
have found the eigenvalues and eigenfunctions by introducing q-deformed
creation and annihilation operators using the complex formalism. It
was shown that how energy eigenvalues of these oscillators in considered
deformation formalism can be derived and also especially we tested
them in limit cases that in both cases ordinary results were recovered.
As well as treatments of energy eigenvalues for real and complex values
of the $q$ parameter were depicted that difference between real and
complex cases were shown clearly. It was seen that for real values
of $q$, in the energy eigenvalues we faced with rapid rise so that
the spectrum got expanded. On the contrary, when we set complex values
for $q$, we were witness that the eigenvalues increased less rapidly
than real case and also the spectrum got compressed that this treatments
resembled us a periodic behaviors.

The connection between our q-deformed Dirac oscillator with quantum
optics is well established via (JC) and (AJC) models, and the existence
of well-known q-deformed version of Zitterbewegung in relativistic
quantum dynamics has been discussed. In the case of very small deformation,
we have calculated the in the pure phase case ($q=e^{i\eta}$) the
partition function and all thermal quantities such as the free energy,
total energy, entropy and specific heat. As application, we have extended
our results to the case of graphene.

\end{document}